\documentclass[preprint,authoryear,5p,twocolumn]{elsarticle} 

\usepackage{amsfonts,amsmath,amscd,amsthm,graphicx,amssymb,subfigure}
\usepackage[utf8]{inputenc} 
\usepackage[table]{xcolor}
\usepackage{multirow}
\usepackage{hhline}
\usepackage{lscape}
\usepackage{ amssymb }

\begin{document}

\title{Cerebral functional connectivity periodically (de)synchronizes with anatomical constraints}

\author[ulg,fn1]{Raphaël Liégeois}
\ead{R.Liegeois@ulg.ac.be}

\author[crc]{Erik Ziegler}
\ead{Erik.Ziegler@ulg.ac.be}

\author[ulg]{Pierre Geurts}
\ead{P.Geurts@ulg.ac.be}

\author[crc,col]{Francisco Gomez}
\ead{fagomezj@gmail.com}

\author[crc]{Mohamed Ali Bahri}
\ead{M.Bahri@ulg.ac.be}

\author[ulg,crc]{Christophe Phillips}
\ead{C.Phillips@ulg.ac.be}

\author[can]{Andrea Soddu}
\ead{	asoddu@uwo.ca}

\author[crc2]{Audrey Vanhaudenhuyse}
\ead{avanhaudenhuyse@chu.ulg.ac.be}

\author[crc]{Steven Laureys}
\ead{	Steven.Laureys@ulg.ac.be}

\author[ulg,Cam]{Rodolphe Sepulchre}
\ead{	R.Sepulchre@eng.cam.ac.uk}

\address[ulg]{Department of Electrical Engineering and Computer Science,
    University of Liège, Liège, Belgium}

\address[crc]{Cyclotron Research Centre, University of Liège, Liège, Belgium}

\address[col]{Computer Science Department, Universidad Central de Colombia, Bogotá, Colombia}

\address[can]{Mind \& Brain Institute, Department of Physics and Astronomy, Western University, London, ON, Canada}

\address[crc2]{Department of Algology and Palliative Care, University Hospital of Liège, Liège, Belgium}

\address[Cam]{Department of Engineering, Trumpington Street, University of Cambridge,  Cambridge CB2 1PZ, United Kingdom}

\fntext[fn1]{Corresponding author}

\begin{abstract}
\vspace{.05cm}
 This paper studies the link between resting-state functional connectivity (FC), measured by the correlations of the fMRI BOLD time courses, and structural connectivity (SC), estimated through fiber tractography. Instead of a static analysis based on the correlation between SC and the FC averaged over the entire fMRI time series, we propose a dynamic analysis, based on the time evolution of the correlation between SC and a suitably windowed FC. 
Assessing the statistical significance of the time series against random phase permutations, our data show a pronounced peak of significance for time window widths around 20-30 TR (40-60 sec). Using the appropriate window width, we show that FC patterns oscillate between phases of high modularity, primarily shaped by anatomy, and phases of low modularity, primarily shaped by inter-network connectivity. Building upon recent results in dynamic FC, this emphasizes the potential role of SC as a transitory architecture between different highly connected resting state FC patterns. Finally, we show that networks implied in consciousness-related processes, such as the default mode network (DMN), contribute more to these brain-level fluctuations compared to other networks, such as the motor or somatosensory networks. This suggests that the fluctuations between FC and SC are capturing mind-wandering effects. 
\end{abstract}


\maketitle

\section*{Introduction}
The human brain shows organized spatio-temporal activity even in task-free or ``resting-state" conditions which is characterized by very slow ($<0.1$Hz) fluctuations of the fMRI Blood Oxygen Level Dependent (BOLD) signal \citep{Gusnard2001,Greicius2003}. Separate and spatially distinct cerebral regions have also been shown to exhibit coherent activity patterns as measured by the correlation between regional fMRI BOLD time series. This measure of so-called functional connectivity (FC) (see \citet{Friston2011} for a review) is organized in robust resting-state networks \citep{Beckmann2005,Damoiseaux2006,Moussa2012}, and has been used to explore a range of properties such as cognition \citep{Richiardi2011,Heine2012}, emotions \citep{Eryilmaz2011}, and learning \citep{Bassett2011}.\\
From an anatomical point of view, structural connectivity (SC) and its multi-scale spatial organization have also been characterized \citep{Sporns2004,Sporns2005} and linked to brain diseases \citep{Kaiser2013,Griffa2013,Engel2013} and genetic influences \citep{Jahanshad2013,Ziegler2013}.\\
The relationship between SC and FC, and more particularly the way cerebral anatomy shapes neuronal functions is a question that has been addressed ever since neuroimaging techniques allowed to collect both structural and functional information \citep[e.g.][]{McIntosh1994}. Different approaches have been used to tackle this question, such as direct comparison of functional and structural connectivities \citep{Kotter2000,Sporns2000}, graph theory \citep{Passingham2002,Bullmore2009}, and model based approaches to explain the link between SC and FC \citep{Koch2002}. However, it is only recently that a clear link between SC and FC \citep{Honey2009,Heuvel2009}[reviewed in \citet{Damoiseaux2009}] has been established, allowing for testable models \citep{Honey2010,Deco2012}.\\
Meanwhile, the classical approach of assuming FC as constant during resting-state recordings \citep{Bullmore2009,Friston2011} has also evolved recently. We will refer to this assumption as a \emph{static} analysis of FC that treats FC as a static quantity, averaging FC over the entire time series. In contrast, many recent studies have emphasized the importance of treating FC as a dynamical quantity, that is, evolving in time \citep{Hutchison2013,Park2013}. Different tools have been proposed to introduce temporal variations into the analyses of FC, such as sliding windows \citep{Sakoglu2010,Bassett2011,Jones2012,Shirer2012,Allen2012,Handwerker2012}, single-volume co-activation patterns (CAPs) \citep{Tagliazucchi2012,Liu2013,Amico2014}, as well as a combination of sliding windows and other methods, such as Independent Component Analysis \citep{Kiviniemi2011} or Principal Component Analysis \citep{Leonardi2013}. For a review of these methods, see \citep{Hutchison2013}.\\
Using a dynamical framework different studies further showed that dynamical FC (dFC) can be seen as the transition between several FC patterns \citep{Gao2010,Deco2013a,Yang2014} presenting different modularity and efficiency properties \citep{Lv2013,Sidlauskaite2014}.  
Finally, many groups have explored day-dreaming, or mind-wandering using functional imaging. The networks implied in these processes are mainly the default mode network (DMN) \citep{Kucyi2014,Fox2013} and the cognitive control network (CCN) \citep{Christoff2009} as well as their interplay \citep{Hasenkamp2012}. The dynamical properties of mind-wandering have also been studied and characteristic frequencies on the order of 0.03-0.05 Hz were found \citep{Bastian2013,Vanhaudenhuyse2011}.\\
In this work we study how the correlation between FC and SC evolves in time using a sliding window approach.  We aimed to demonstrate that considering the dynamics of the FC-SC correlation can reveal some information that is hidden in a classical static analysis \citep{Honey2009,Deco2012,Deco2013}. The first part of the paper addresses the issue of selecting a proper time window \citep{Allen2012,Hutchison2013}. Using the selected time window, the second part explores the dynamic interactions between FC and SC at the brain level and in particular subnetworks such as the default mode and cognitive control networks.\\

\section*{Material and methods}

\subsection*{Participants}
Data was collected from 14 healthy volunteers (age range 45$\pm$7 years, 7 women, all right-handed). Volunteers gave their written
informed consent to participate in the study, which was approved by the Ethics Committee of the Faculty of Medicine of the University of Liège.

\subsection*{Diffusion Weighted Imaging}

\textbf{DWI acquisition} 
Data was acquired on a 3T head-only scanner (Magnetom Allegra, Siemens Medical Solutions, Erlangen, Germany) operated with the standard transmit-receive quadrature head coil. A high-resolution T1-weighted image was acquired for each subject (3D magnetization-prepared rapid gradient echo sequence, field of view = 256$\times$240$\times$120 mm$^3$, voxel size=1$\times$1$\times$1.2mm). A single unweighted (b = 0) volume was acquired followed by a set of diffusion-weighted (b = 1000) images using 64 non-colinear directional gradients. This sequence was repeated twice for a total of 130 volumes.

\textbf{Processing}

The processing pipeline was developed in Nipype \citep{Gorgolewski2011} and has been described in more detail previously \citep{Ziegler2013}. Structural MR images were first segmented using the automated labeling of Freesurfer \citep{Desikan2006}. Segmented structural images were then further parcellated using the Lausanne2008 atlas for a total of 1015 regions of interest (ROIs) \citep{Cammoun2012}. Diffusion-weighted images were aligned using FSL to the initial unweighted volume to correct for image distortions arising from eddy currents \citep{Smith2004}. Fractional anisotropy maps were generated, and a small number of single-fiber (high FA) voxels were used to estimate the spherical harmonic coefficients of the response function from the diffusion-weighted images \citep{Tournier2004,Tournier2007}. Using non-negativity constrained spherical deconvolution, fiber orientation distribution (FOD) functions were obtained at each voxel. For our dataset with 64 directions, we used the maximum allowable harmonic order of 8 for both the response estimation and spherical deconvolution steps. Probabilistic tractography was performed throughout the whole brain using seeds from subject-specific white-matter masks and a predefined number of tracts.\\
Fiber tracking settings were as follows: number of tracks = 300,000, FOD amplitude cutoff for terminating tracks = 0.1, minimum track length = 10 mm, maximum track length = 200 mm, minimum radius of curvature = 1 mm, tracking algorithm step size = 0.2 mm.\\
Using tools from Dipy (Diffusion in Python, http://nipy.sourceforge.net/dipy/), the tracks were affine-transformed into the subject's structural space and connectome mapping was performed by considering every contact point between each tract and the outlined regions of interest \citep{Ziegler2013}.
 \begin{figure*}[t!]
	\centering
		\includegraphics[width=1\textwidth]{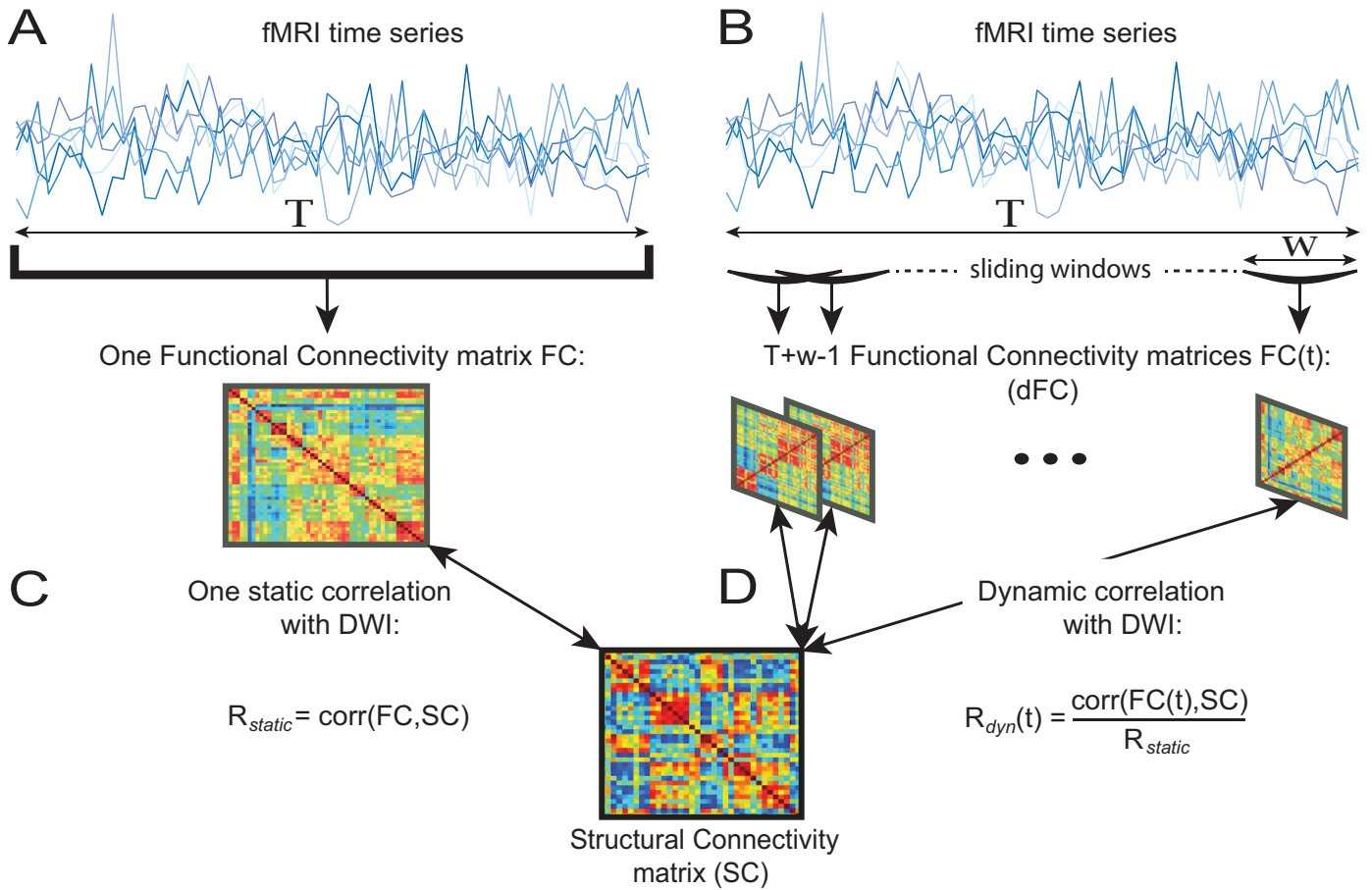}
	\caption{Comparison between the static and the dynamic analysis of the correlation between structural and functional connectivities (SC and FC, resp.). {(A)} FC is computed using the whole fMRI time course. {(B)} dFC is computed in windows of the fMRI time courses that are slid across the whole fMRI time course. {(C)} Static correlation $R_{static}$ between SC and FC, as computed in e.g. \citet{Honey2009}. {(D)} Dynamic correlation $R_{dyn}(t)$ between SC and FC(t), normalized by $R_{static}$ and used in the present work.}
	\label{fig:Methods}
\end{figure*} 

\subsection*{Functional data}

\textbf{BOLD acquisition}
Three hundred multi-slice T2*-weighted functional images were acquired with a gradient-echo echo-planar imaging sequence using axial slice orientation and covering the whole brain (32 slices; voxel size: 3$\times$3$\times$3mm$^3$; matrix size 64$\times$64$\times$32; repetition time = 2000 ms; echo time = 30 ms; flip angle = 78º; field of view = 192$\times$192 mm$^2$). The three initial volumes were discarded to avoid T1 saturation effects. For anatomical reference, a high-resolution T1-weighted image was acquired for each subject. 

\textbf{Processing} 
fMRI data preprocessing was performed using Statistical Parametric Mapping 8 (SPM8; www.fil.ion.ucl.ac.uk/spm). Preprocessing steps included slice-time correction, realignment and adjustment for movement-related effects, coregistration of functional onto structural data, segmentation of structural data, spatial normalization into standard stereotactic Montreal Neurological Institute (MNI) space, and spatial smoothing with a Gaussian kernel of 8 mm full width at half-maximum. Further motion correction was applied using ArtRepair toolbox for SPM\footnote{http://cibsr.stanford.edu/tools/human-brain-project/artrepair-software.html} which corrects for small, large and rapid motions, noise spikes, and spontaneous deep breaths. Finally, linear regression of mean global BOLD signal, mean ventricular BOLD signal and mean white matter BOLD signals from each voxel was performed. Even if it is still a debated question it could be argued that global signal regression \citep{Macey2004} could induce spurious correlations in our analysis \citep[e.g.][]{Murphy2009}. However, it has been shown in \citet{Honey2009} that global signal regression is an essential step in order to reveal the correlation between structural and functional connectivities. Since this anatomy-function link is the main focus of our paper we regressed out the global signal.

The timecourse for each region-of-interest was extracted by taking the average signal over all voxels in each ROI defined following the same parcellation procedure as for anatomical data.
 \begin{figure*}[t!]
	\centering
		\includegraphics[width=1\textwidth]{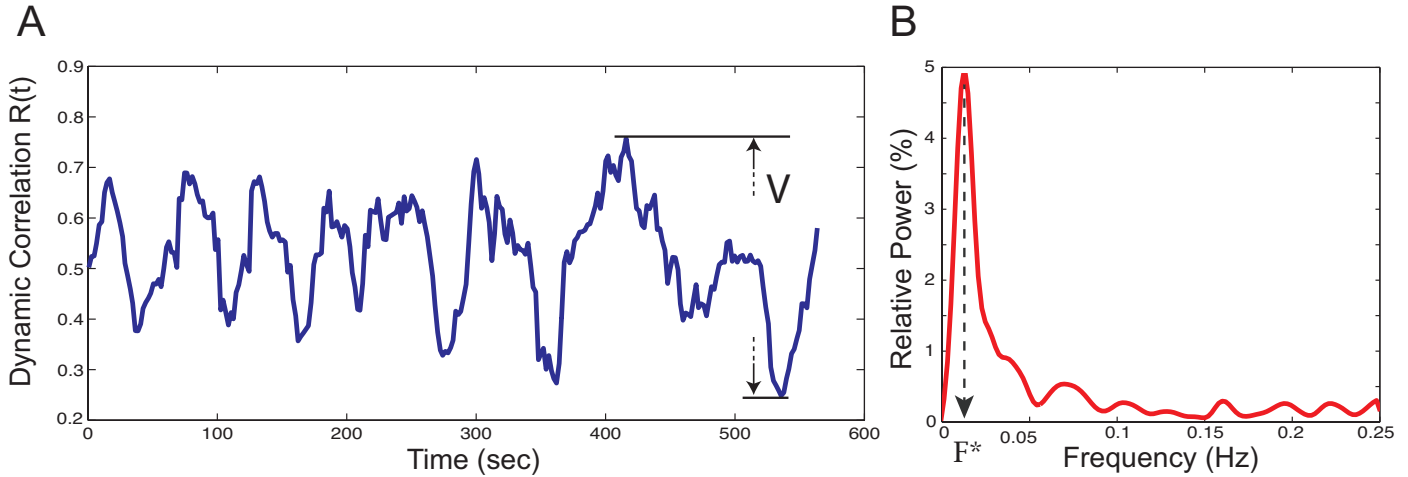}
	\caption{(A) Temporal evolution of the \emph{dynamic correlation} between SC and FC. (B) Corresponding power spectrum. Results are shown for a representative subject and a window width of $w=13$TR. The static correlation = 39.1\%, $V=51.7\%$ and $F^*=0.012$Hz}
	\label{fig:Subj7}
\end{figure*} 

\subsection*{Sliding Window for FC Analysis}

In order to explore the dynamics in the correlations between structural and functional connectivities we repeated the computation of the FC matrices from truncated portions of the fMRI time series, as previously presented \citep{Chang2010,Hutchison2012,Allen2012,Leonardi2013}.\\
Denoting by $T$ the number of volumes in the fMRI time series and considering a window width $w$ (the width of the truncated portions), we computed $T-w+1$ successive FC matrices from the truncated fMRI time series in each particular window, each one being shifted forward by one TR with respect to the previous one (Figure \ref{fig:Methods}B).\\
We used window widths ranging from $2$ to $100$ volumes, corresponding to $4$ to $200$ seconds, to explore the dynamics between structural and functional connectivities.

\subsection*{Dynamic correlation between structural and functional matrices}

We then computed the correlation between all the FC matrices and the SC matrix (Figure \ref{fig:Methods}D). This included log-rescaling of the non-zero values in the SC matrix such that the range in both connectivity matrices have the same order of magnitude (see Supplementary Material for further details).\\
The evolution of this correlation was normalized by the \emph{static correlation} between the SC and the FC matrices computed using the whole fMRI time series (Figure \ref{fig:Methods}C,D), resulting in what we call the \emph{dynamic correlation}, denoted by R(t).\\
In order to characterize the fluctuations of the dynamic correlation the power spectrum of R(t) was computed using Welch's method \citep{Welch1967} and normalized such that $\int_0^{0.25}P(f)dF=1$ where $P(f)$ is the power spectral content corresponding to frequency $f$. 

\subsection*{Values of interest}
In order to characterize dynamics observed in the time-evolving \emph{dynamic correlation} curves and their corresponding spectral power, we used two markers:

\begin{itemize}
\item V is the range of variation in the \emph{dynamic correlation}, computed as the difference between its maximal and minimal value, in percent. V is used to highlight the phases of (de)synchronization between SC and FC (Fig \ref{fig:Subj7}A),
\item $F^*$ is the frequency of maximal relative spectral power (in Hz), and corresponds to the main oscillatory mode of a time course, such as in Figure \ref{fig:Subj7}B.
\end{itemize}

\subsection*{Statistical significance of the observed dynamics}

A main issue in fMRI time series analyses is to disentangle the neuronal dynamics from noise \citep{Handwerker2012}. To this end we performed the same computations as the one described in Figure \ref{fig:Methods} using surrogate data obtained by \emph{phase randomization} in the Fourier domain of the fMRI volumes \citep{Theiler1992}, similar to what is presented in \citet{Allen2012}, for example (see Supplementary Material for details). Doing so leaves the \emph{static correlation} unchanged because the overall covariance structure is preserved, whereas the evolution of the dynamic correlation R(t) using windowing will be totally rearranged. 

We observed larger fluctuations (higher $V$) of R(t) in the original data compared to the surrogate data. Hence, we chose this marker to test for differences between the results obtained with ordered and phase randomized fMRI time series. For each value of window width and each subject we did 1000 permutations \citep[see e.g. Chap 3.5 in][]{Edgington1969} and computed the z-score corresponding to the following null hypothesis:

$$\mathcal{H}_0=\{V_{ord} \ngtr V_{rand}\}$$

\noindent where $V_{ord}$ (resp. $V_{rand}$) is the range of variation of R(t) in the original ordered (resp. surrogate) data. \\

\begin{figure}[h!]
	\centering
		\includegraphics[width=0.5\textwidth]{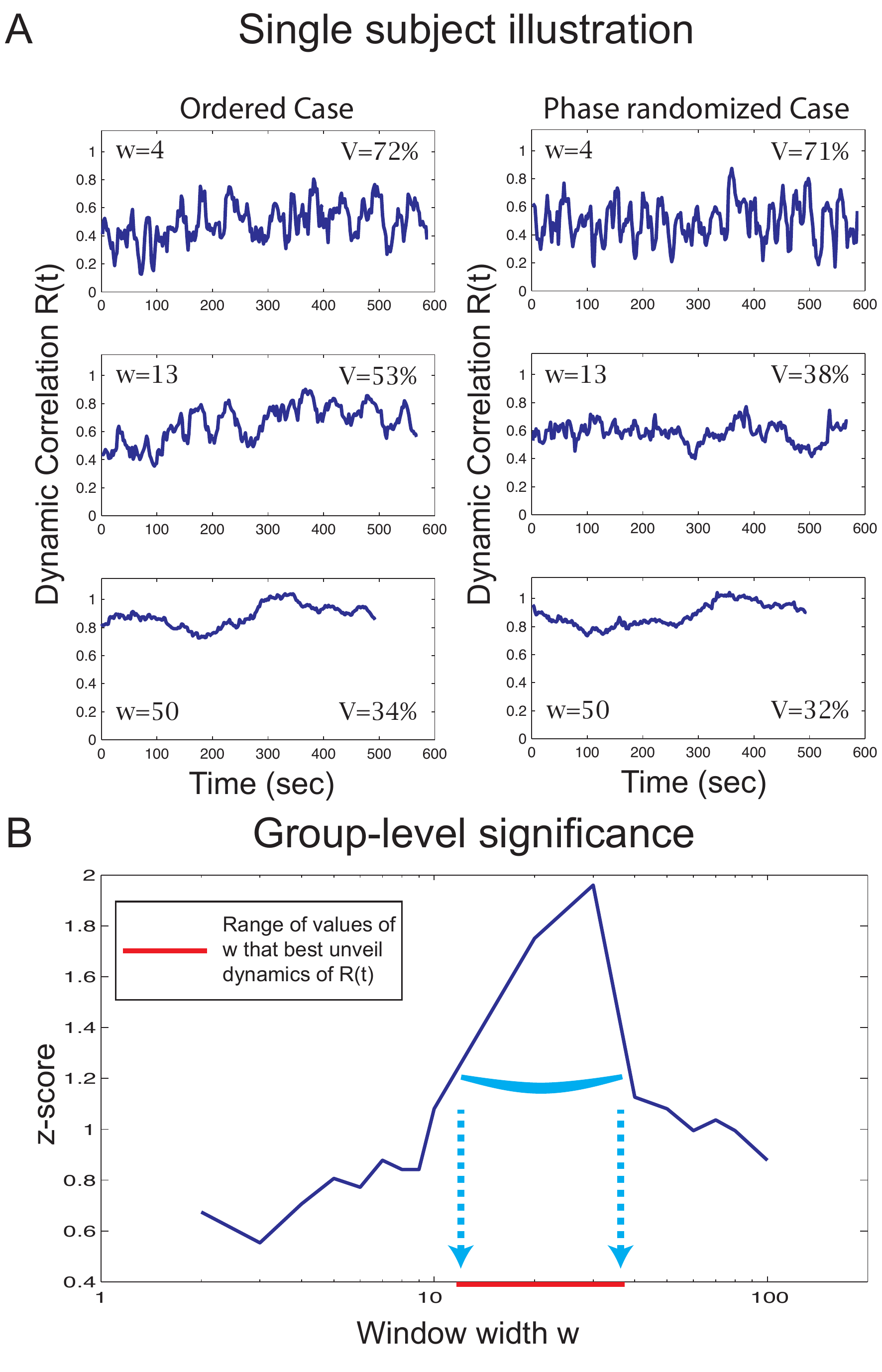}
	\caption{(A) R(t) for different window widths for a representative subject in the original dataset and one sample of the surrogate dataset. (B) Estimation of the statistical significance region at the group level based on the range of variation V of R(t)}
	\label{fig:stat_sign}
\end{figure}

The group level significance curve presented in Figure \ref{fig:stat_sign}B was computed from the z-scores of all the subjects. This technique is known as the Stouffer's method \citep{Stouffer1949} , and is detailed in the Supplementary Material.

\subsection*{Graph theory metrics}

In order to further characterize FC during the phases of (de)synchronization with SC, we used three common graph metrics of FC considered as a weighted undirected graph \citep{Bullmore2009}. In this context, each ROI is considered as a \emph{node} of the graph and the level of correlation between each two regions $i$ and $j$ $FC_{i,j}$ is the weight of the \emph{edge} connecting these two regions. Since FC is symmetric, it follows that the corresponding graph is undirected. We used the three following metrics on the whole FC matrices:

\begin{itemize}
\item \emph{Density} is the number of total connections divided by the number of possible connections \citep{Sporns2002},
\item \emph{Efficiency} measures how 'close' every two nodes are in the graph. It is inversely related to the characteristic path length \citep{Onnela2005,Rubinov2010},
\item \emph{Modularity} quantifies to which degree a network can be subdivided into distinct groups \citep{Newman2004}.
\end{itemize}

The Brain Connectivity Toolbox \citep{Rubinov2010} was used to evaluate the value of these three markers during phases of high and low correlation between $SC$ and $FC(t)$. For each subject, averaged top and bottom 5\% of FC(t) matrices were selected, sorted by R(t) value (see Figure \ref{fig:wwdfg} for more details). Since density was designed for binary graphs, we binarized the FC matrices (only for this marker) using a 0.1 threshold. It should be noted that the choice of the threshold does not influence the trend observed in Figure \ref{fig:graph} \emph{left}. 

\subsection*{Network-level analysis}
All computations above have been performed at the whole-brain level. In order to test the hypothesis that the fluctuations in R(t) can be linked to consciousness-related processes, we repeated the procedure presented in Section 2.7 in selected regions, defined by the Lausanne Atlas \citep{Hagmann2008}. We tested the fluctuations in the precuneus for the default mode network (DMN), and the supramarginal region for the cognitive control network (CCN). These two regions are known to be involved in mind-wandering processes \citep[e.g.][]{Fox2013}. We also tested these fluctuations in two regions of the somato-motor network: the precentral gyrus for the motor network (MOT) and in the postcentral gyrus for the somatosensory network (SS).\\
In order to test the similarity between the effect observed at the brain level and in the four networks, we computed the correlation between the z-scores equivalent to the curves presented in Figures \ref{fig:stat_sign}B and \ref{fig:nets}. This correlation coefficient is indicated next to the corresponding curve in Figure \ref{fig:nets}.\\
Next, in order to test the difference between the effects observed in the four networks and for w = 20 TR, we performed a paired t-test using the equivalent z-scores obtained at the first step of Stouffer's method (see Supplementary Material) for w = 20 TR. The results are represented in the table on the left of Figure \ref{fig:nets}.\\

\section*{Results}
\subsection*{Statistical significance of dynamical correlation}

The dynamical correlation for a representative subject is shown in Figure \ref{fig:Subj7} for a window width = 13 TR. In this example the dynamic correlation varies from 24\% to 76\% ($R=52\%$) of the static correlation, and the peak spectral power is $F^* = 0.012$Hz.\\

The choice of window width $w$ affects the way dynamical correlation is captured (see Figure \ref{fig:ww} in Supplementary Material). Significance of observed fluctuations as a function of $w$ is represented in Figure \ref{fig:stat_sign} and was tested by comparison against phase randomized fMRI time series as explained in the Methods section.\\

Figure \ref{fig:stat_sign}A illustrates the fact that the difference between ordered and phase randomized fMRI time series as captured by V is more pronounced for intermediate values of $w$. At the group level, a peak of statistical significance can be observed around $w=20$ TR (Figure \ref{fig:stat_sign}B) hence this is the window width that we will use in the following analyses.

\subsection*{Phases of (de)synchronization between functional and structural connectivities}
\begin{figure*}[t]
	\centering
		\includegraphics[width=1\textwidth]{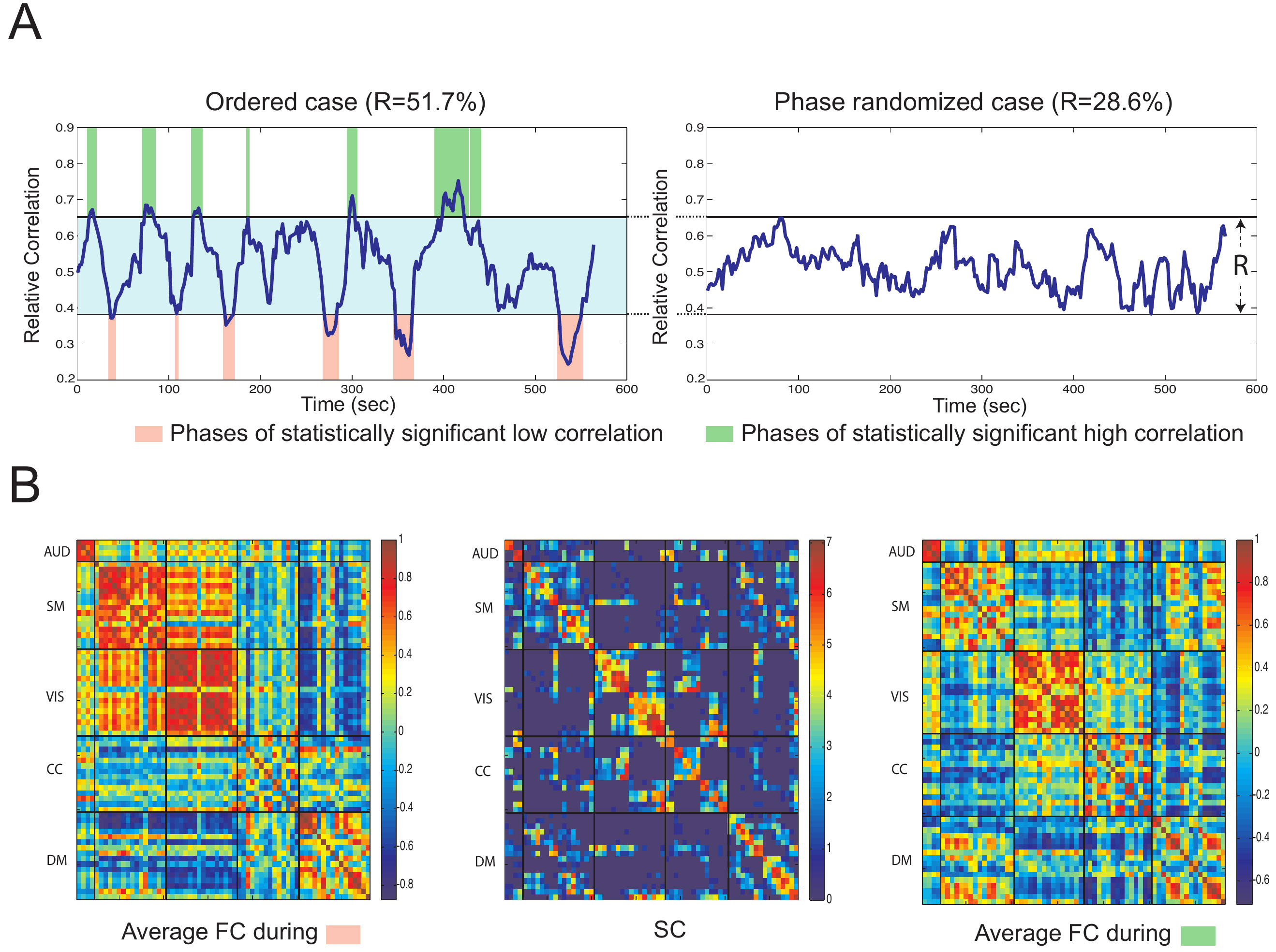}
	\caption{(A) Illustration of the phases of (de)synchronization between FC and SC for a representative subject based on the difference in the values of V in both cases. (B) \emph{left} Average FC matrix computed by averaging the FC matrices that have the 5\% lowest correlations with SC. \emph{middle} Structural connectivity matrix. \emph{right} Average FC matrix computed by averaging the FC matrices that have the 5\% highest correlations with SC.}
	\label{fig:wwdfg}
\end{figure*}

We show in Figure \ref{fig:wwdfg}A the statistically significant phases of (de)synchronization between FC(t) and SC, corresponding to the highest and lowest values of R(t), respectively. The average patterns of FC(t) computed during these phases are represented in Figure \ref{fig:wwdfg}B as well as the constant structural connectivity matrix, for one particular subject.

\begin{figure}[h!]
	\centering
		\includegraphics[width=0.5\textwidth]{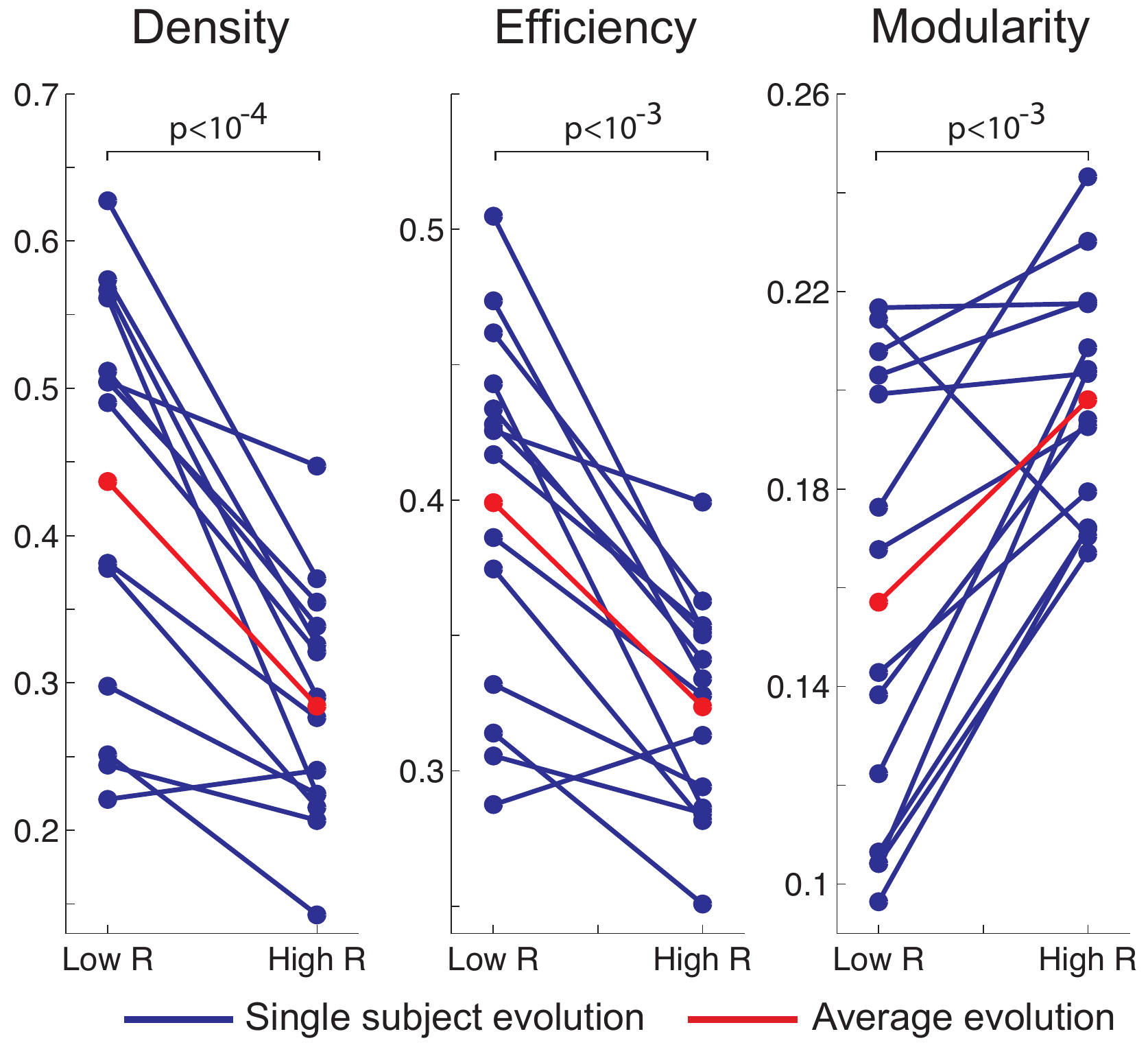}
	\caption{Density, Efficiency and Modularity of FC averaged over the 5\% lowest correlations with SC (low R - left columns) and FC averaged over the 5\% highest correlations with SC (high R - right columns) for all the subjects. The group mean is represented in red.}
	\label{fig:graph}
\end{figure}

Density, efficiency and modularity of the high and low FC patterns for all the subjects are represented in Figure \ref{fig:graph}.

Density and efficiency appear to be significantly lower ($p<10^{-4}$ and $p<10^{-3}$ using a paired t-test) when the correlation between FC(t) and SC is high whereas modularity increases at the same time ($p<10^{-3}$).
\subsection*{Networks implied in the fluctuations}

\begin{figure}[h!]
	\centering
		\includegraphics[width=0.5\textwidth]{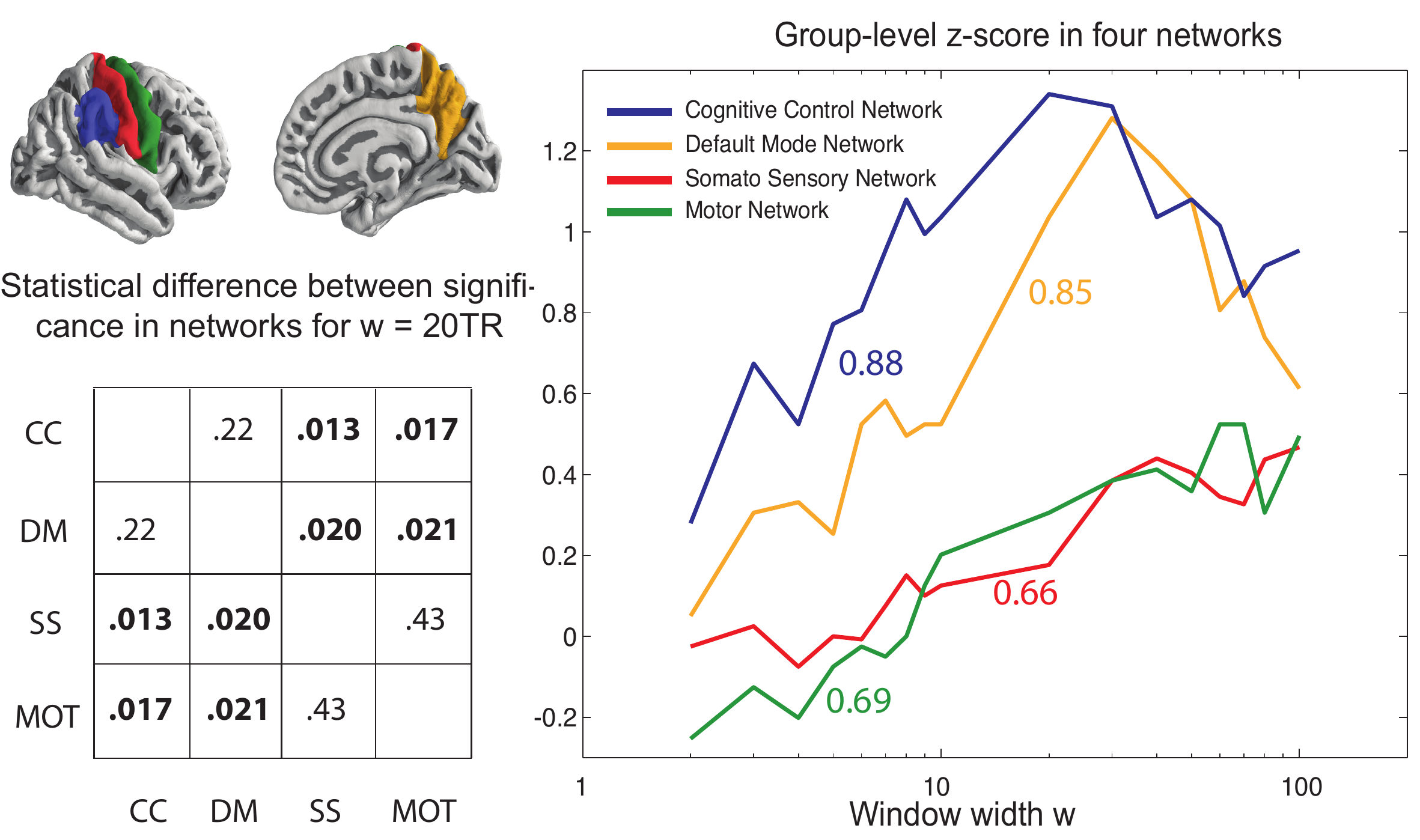}
	\caption{ \emph{Right} Statistical significance of fluctuations for four different regions in four different networks. DMN: precuneus; CC: supramarginal; SS: postcentral; MOT: precentral. Correlation coefficients between the equivalent z-scores curves of the four networks and the brain-level curve presented in Figure \ref{fig:stat_sign}B are represented on the curves. \emph{Left} Inter-network difference of significance tests deduced from the equivalent z-scores and for w=20 TR.}
	\label{fig:nets}
\end{figure}

The statistical significance of the fluctuations in four different regions of the brain, representative of four networks, are represented in Figure \ref{fig:nets}.  

The characteristic 'V' shape observed at the whole-brain level in Figure \ref{fig:stat_sign}B is present in the DMN and the CC but not in the MOT and SS networks, as indicated by the correlation coefficients between the equivalent z-scores curves (see Methods section). In addition, the significance of the fluctuations appears to be lower in the DMN and the CCN compared to the SS and MOT networks as shown by the p-values in the table on the left of Figure \ref{fig:nets}.

\section*{Discussion}

\subsection*{A) Significance of fluctuations observed using the sliding window approach}

\begin{figure}[h!]
	\centering
		\includegraphics[width=0.5\textwidth]{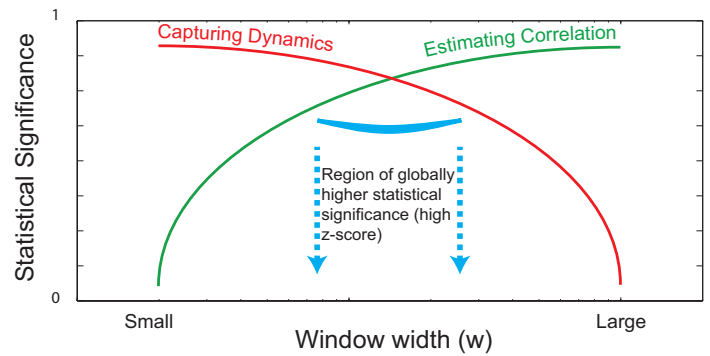}
	\caption{Interpretation of Figure \ref{fig:stat_sign}: tradeoff between capturing dynamics and estimating correlation}
	\label{fig:inter}
\end{figure}

Sliding window techniques have been widely used in recent studies in order to analyze FC dynamics. \citet{Allen2012} used a width of 22 TR (TR=2 sec) to track oscillations in FC dynamics, \citet{Shirer2012} showed that considering a width above 15-30 TR (TR=2 sec) allows for robust estimates of the FC without considering dynamics. More recently, \citet{Leonardi2013} used widths ranging from 20 to 120 TR (TR=1.1 sec) and observed different ``eigenconnectivity patterns" depending on the window that is used and \citet{Hutchison2012} also found different results with window width going from 10 to 120 TR (TR=2 sec).\\ 
Our study reveals a peak of statistical significance in the observed fluctuations around $w=20-30$ TR (TR=2 sec). This peak is observed using the \emph{range of variation V} of R(t). Another statistical test based on the variance of the R(t) curves (see Figure \ref{fig:varpv} in Supplementary Material) also shows a peak for values of $w$ around 20 TR which reinforces our conclusions about which window width should be used.\\
Our results are consistent with a general tradeoff in time series analyses: longer windows improve the estimation of the correlation but mask the dynamics because they act as low-pass filters. This effect is illustrated by the green curve in Figure \ref{fig:inter} and was studied by \citet{Shirer2012}.

Our analysis provides the additional insight that considering higher values of $w$ does not capture the FC neuronal dynamics, which is illustrated by the red curve in Figure \ref{fig:inter}. Indeed, we can compare windowing of correlations to a moving average which is a low-pass filter with cutoff frequency $f_c$ that decreases when $w$ increases \citep[e.g.][Chap.15]{Smith1997}.\\
We hereby also want to stress the importance of assessing the neuronal origin of fluctuations that are observed in dynamic functional connectivity \citep{Hutchison2013}. The simple significance testing framework proposed in the present work is believed to be an important prerequisite to further interpretation of observed functional connectivity fluctuations.

\subsubsection*{Limitations}
The sliding window acts as a low-pass filter. In our case, considering $w=20$ TR $= 40$ sec results in a cutoff frequency $f_c \approx 0.02$ Hz. Hence, a robust estimation of FC, which requires a window width $w\approx 20$ TR, necessarily filters out the FC dynamics happening at higher frequencies than $\approx 0.02$ Hz.\\
This limitation should be taken into account when interpreting results of dynamical FC analyses using sliding windows. This is also a call for more advanced identification methods that could better estimate the FC and push the green curve of Figure \ref{fig:inter} to the left, consequently freeing the access to higher frequency dynamics.

\subsection*{B) Phases of (de)synchronization between functional and structural connectivities}
The link between structural and functional connectivities was established a few years ago \citep{Honey2009,Heuvel2009}. Thereafter, a lot of interest has been devoted to deepen the understanding of how anatomical constraints shape functional connectivity \citep{Honey2010,Breakspear2010,Cabral2011,Deco2012}, and how this relationship can be affected by different pathologies \citep{Kwaasteniet2013,Schouwenburg2013}.\\
In most of these studies either the dynamics of FC are not taken into account, or it is modeled, but the information coming from the data and used to assess models is deduced with a static approach of FC \citep[e.g.][]{Deco2013}.\\
To our knowledge, our paper is the first data driven attempt to study the dynamical relationship between SC and FC. More specifically, we show in Figure \ref{fig:stat_sign} that there are statistically significant (i.e. resulting from the neuronal dynamics, not noise) phases of (de)synchronization between the functional correlation and the anatomical constraints. When using statistically significant values of $w$ such as $w=20$ TR, the range of variation R is on the order of 52\% of the static correlation, compared to 34\% in the randomized case, meaning that the correlation between FC and SC is significantly high at some points and significantly low at some other points.

\subsubsection*{Structure as a switch between functional brain states}

Diverse studies have recently highlighted the presence of different and successive functional connectivity states, even at rest \citep{Lv2013,Gao2010,Deco2013a,Yang2014,Sidlauskaite2014}. The results shown in Figures \ref{fig:wwdfg} and \ref{fig:graph} suggest that the dynamic reorganization of functional connectivity patterns is shaped by anatomy. More particularly it can be observed from Figure \ref{fig:graph} that phases of high correlation between FC and SC correspond to functional connectivity patterns that have poor efficiency and high modularity. The interpretation is that during these phases the brain is poorly functionally connected, and organized in modules shaped by anatomy with few inter-modules connections \citep{Newman2004}. On the other hand, during phases of low correlation between FC and SC, the number of inter-modules functional connections increases, resulting in highly connected FC patterns. \\
Very recently, \citet{Messe2014} evinced a decoupling between anatomy-defined networks and other networks resulting from stationary and non-stationary FC dynamics, but not related to anatomy. Combined to our results, these observations lead us to propose that anatomy could periodically play the role of a relay that guides switches between different highly connected FC patterns not shaped by anatomy (red regions in Figure \ref{fig:wwdfg}A), alternating with phases of lower efficiency and higher modularity, defined by SC architecture (green regions in Figure \ref{fig:wwdfg}A). This interpretation echoes another recent work \citep{Zalesky2014} in which the most dynamic connections are shown to be inter-modular, and support the emergence of temporary phases of high functional efficiency.\\

\subsubsection*{R(t) as a foot-print of mind-wandering}

The strength of the link between anatomy and functional correlates has been shown to vary between different attentional states \citep{Baria2013}. Several of our results provide temporal and spatial indicators of the statistical correlation between R(t) and \emph{mind-wandering} processes. First, we showed that the fluctuations of R(t) are statistically significant when a window width in the range $20-30$ TR is used, corresponding to a main oscillatory mode $F^*$ on the order of 0.01 $\pm$ 0.003 Hz (Figure \ref{fig:ww}C in Supplementary Material), which is quite close to the typical frequencies of networks mediating consciousness markers such as awareness of environment and of self \citep{Vanhaudenhuyse2011}. Second, the results presented in Figure \ref{fig:nets} show that the significance curves for DM and CC present the characteristic 'V' shape that is found at the brain level, whereas the curves in the MOT and SS networks are almost flat resulting in lower correlation coefficients for these networks. In addition, the level of significance appears to be significantly higher in the DMN and the CC compared to the other two networks (Figure \ref{fig:nets}(left), tested for w=20 TR). This suggests that the fluctuations that are observed in R(t) are driven by neuronal dynamics happening in the DM and CC networks and not in the MOT and SS networks and are consequently at least partly correlated to mind-wandering effects. \\
Going further, \citet{Doucet2012} reported that mind-wandering was correlated with fluctuations of functional modular organization, inner-oriented activities being associated to phases of low inter-modular connectivity. This is an additional evidence supporting the interpretation of R(t) as reflecting mind-wandering processes.

\subsubsection*{Limitations}

We use $F^*$ to characterize the fluctuations of R(t). It is interesting to note that we were not able to distinguish ordered from phase randomized time courses using $F^*$ (see Figure \ref{fig:varf} in the Supplementary Material), suggesting that $F^*$ is imposed by the sliding window method and by $w$ (Figure \ref{fig:ww}C in the Supplementary Material) and is not \emph{a priori} capturing neuronal dynamics. Hence it is not surprising to find similar values of $F^*$ in studies using a similar window width: \citet{Allen2012} found oscillations at 0.005-0.015 Hz using a 22 TR (44 sec) windowing. However, as argued in \citet{Hutchison2013}, this does not imply that the value of $F^*$ for ordered fMRI time series has a non-neuronal origin. It is used here because the significance of oscillations was independently assessed using two other markers of R(t).

Finally it should be highlighted that having a frequency peak at around 0.01 Hz for the values of w that are statistically significant does not mean that the dynamics are only occurring at these frequencies. As explained earlier, hypothetical dynamics happening at higher frequencies are filtered out when we use 20 TR windowing and it is, for example, difficult to assess the correspondence with the consciousness markers from a dynamical point of view because they are happening at higher frequencies.

\subsection*{C) Future work} 

The results of this paper call for several methodological refinements. We have presented a hypothesis-driven approach in order to test which network contributes most strongly to the brain-level oscillations. It would be interesting to use data-driven approaches in order to finely identify which areas contribute more strongly to the overall fluctuations of R(t). Addressing this problem requires a solution for the dimensionality issues raised in this application. Indeed, if one wants to derive networks without needing any a priori knowledge, we would have either to use adequate downscaling or to consider every pair of voxels as a variable, that is $\mathcal{O}(N^2)$ variables where $N$ is the number of ROIs, 1015 in our case.\\
Moreover, windowing is an approach that shows some limitations, and future studies should consider more advanced techniques for the dynamics identification. These alternatives may allow for clearer identification of dynamics of functional connectivity, and could help unveil processes occurring at higher frequencies such as mind-wandering. As fast scanning becomes feasible with new scanners and parallel imaging, one simple way to test this hypothesis would be to use smaller TRs, up to 1 sec, in order to increase the lowpass cutoff frequency of the windowing process.\\
It could also be worth completing the present multi-modal analysis with other imaging modalities such as EEG. It has, for example, been shown that EEG micro-states can be considered as building blocks of cognition and shape fMRI networks \citep{VandeVille2010}. Hence, including the high-resolution temporal information provided by EEG measurements could lead to a better understanding of the interaction between anatomy and function and its interpretation in terms of cognitive processes. 

\section*{Conclusions}

The contribution of the present paper is twofold. From a methodological point of view we highlight some characteristics of the sliding window technique to reveal functional connectivity dynamics. Our results suggest that the width of those windows should be chosen around the 20-30 TR (40-60 sec) range to both provide a robust estimate of the correlation and capture significant functional connectivity neuronal dynamics. For smaller or higher values, we could not significantly distinguish functional connectivity dynamics from noise with similar properties.\\
Next, we use a suitable window width to show that that dynamical functional connectivity oscillates between states of high modularity, mostly shaped by structural connectivity architecture, and states of low modularity, not defined by structural connectivity, during which more inter-network connections take place.\\
Finally, considering that these fluctuations are occuring at a characteristic frequency of $\approx 0.01$ Hz and that the Default Mode and the Cognitive Control networks are highly contributing to their dynamics, we propose that the dynamical correlation between functional connectivity and anatomy is an interesting marker of mind-wandering effects.

\bibliographystyle{my_model2-names}	

\bibliography{Raphael_Liegeois.bbl}

\newpage
\section*{Supplementary Material}
\subsection*{Correlation between SC and FC}

\subsubsection*{Log-Rescaling of SC matrices}
The FC matrices take their values in the [-1,1] interval. On the other hand, the SC matrices that encode the density of fiber tracks between every pair of cortical regions contains exponentially distributed values ranging from $0$ to $10^6$. In order to improve the match between these two ranges of variation and consequently increase the values of $R_{stat}$, we considered the logarithm of every non-zero element in the SC matrices, as presented in \citep{Honey2009}. \\
\subsubsection*{Correlation of correlation}
As already mentioned, we are computing correlations between SC and FC which itself is encoding correlations between fMRI time series. It could be argued that this is not appropriate because the variance of the correlation coefficients is not constant on the [-1,1] interval. This could be addressed by applying a variance-stabilizing transformation such as the Fischer transformation but we did not use this approach in the present work for two reasons. First because SC is not encoding a correlation coefficient and hence the matching between the values in FC and SC should not a priori be addressed using this type of transformation, the log-rescaling being an alternative. The other reason is that we wanted to stick to the methodology presented in \citet{Honey2009} in order to explore which additional information we get going from the static to the dynamic case, all other things being equal. Let us finally note that we computed the correlation between FC and SC and did not find significant differences. One possible explanation for that could be that there are few 'extreme' values in FC and hence the Fisher transform does not modify significantly the values in FC since the transformation has very mild effect for correlation coefficients in the [-0.7, 0.7] interval.

\subsection*{Impact of window width}

\begin{figure}[h!]
	\centering
		\includegraphics[width=0.5\textwidth]{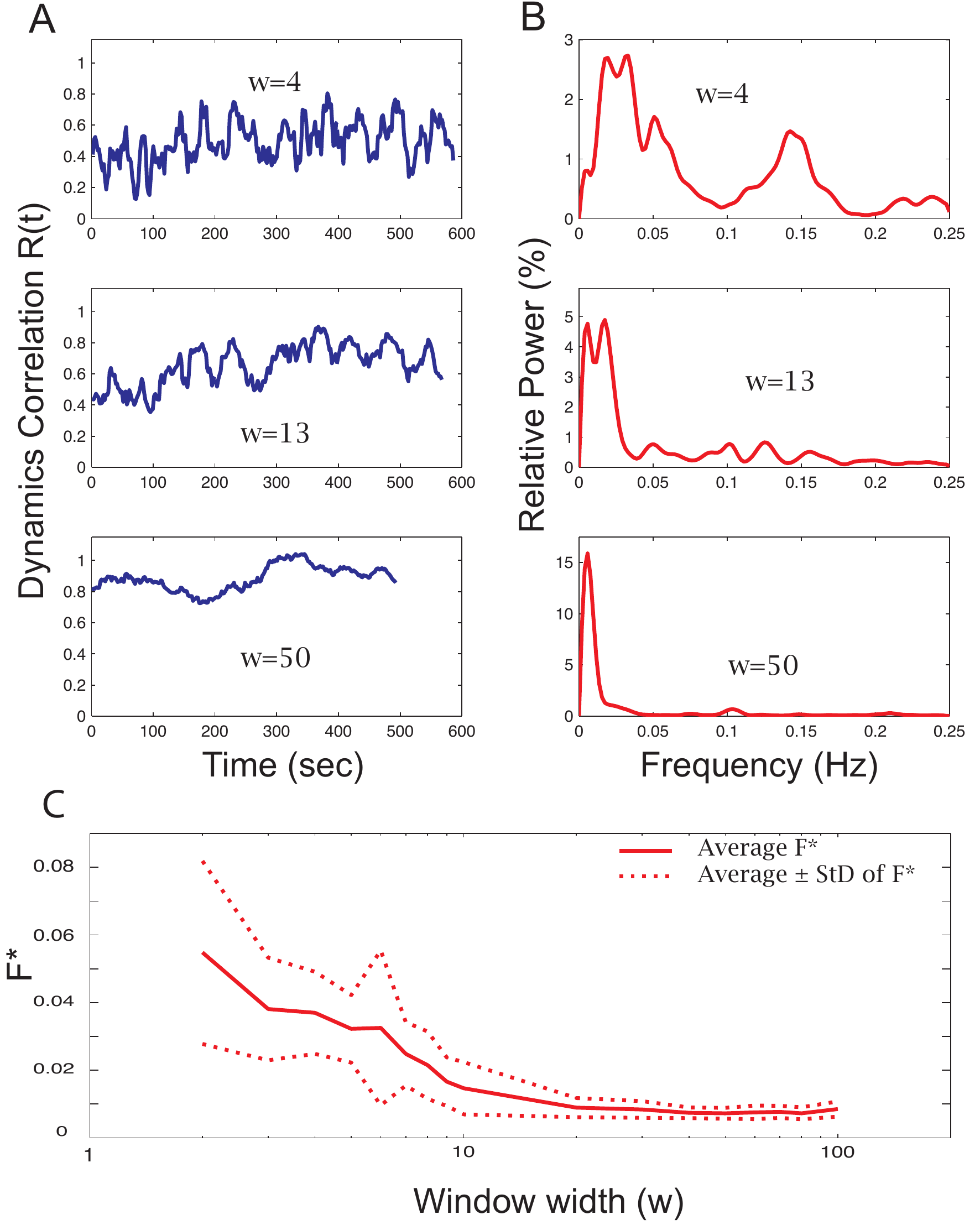}
	\caption{(A) R(t) and (B) corresponding power spectrum for different window widths $w$ and for a representative subject. (C) Mean and standard deviation of $F^*$ for all subjects as a function of the window width $w$}
	\label{fig:ww}
\end{figure}

The impact of the window width $w$ can be observed in Figure \ref{fig:ww} for selected values of $w$.\\ 

We observe that increasing $w$ results in smoothening of the dynamic correlation curve (Figure \ref{fig:ww}A). This can also be observed in the power spectrum, which is globally shifted towards low frequencies when $w$ increases (Figure \ref{fig:ww}B), resulting in a decrease of $F^*$ for higher values of $w$ (Figure \ref{fig:ww}C).\\

\subsection*{Phase randomization of the fMRI time series}
Phase randomization was performed by adding the same random sequence of phases to all the fMRI time series phase spectra. Doing so preserves the mean, the variance and the temporal autocorrelation of the time series but destroys the information contained in the ordering of the fMRI volumes. Comparing original and surrogate datasets then allows to test whether the fluctuations of R(t) are due to noise or result from neuronal dynamics.

\subsection*{Stouffer's method implementation}
Stouffer's method was used to test the following $\mathcal{H}_0$ hypothesis at the group level:
$$\mathcal{H}_0=\{V_{ord} \ngtr V_{rand}\}$$
where $V_{ord}$ (resp. $V_{rand}$) is the range of variation of R(t) in the original ordered (resp. surrogate) data.

It allows for the combination of z-scores from several independent tests bearing upon the same group-level hypothesis. In our case, we proceeded as follows:

\begin{enumerate}
\item Compute the z-score corresponding to the null hypothesis for each subject and each window, denoted by $Z_i, i \in \{1...14\}$.
\item Compute the z-score for each window width at the group-level analysis $Z_G$ as follows:\\
$$Z_G=\frac{\sum_{i=1}^N Z_i}{\sqrt{N}}$$
where N is the number of subjects.
\end{enumerate}

\subsection*{Determining the statistical significance using other markers}
\begin{figure}[h!]
	\centering
		\includegraphics[width=0.5\textwidth]{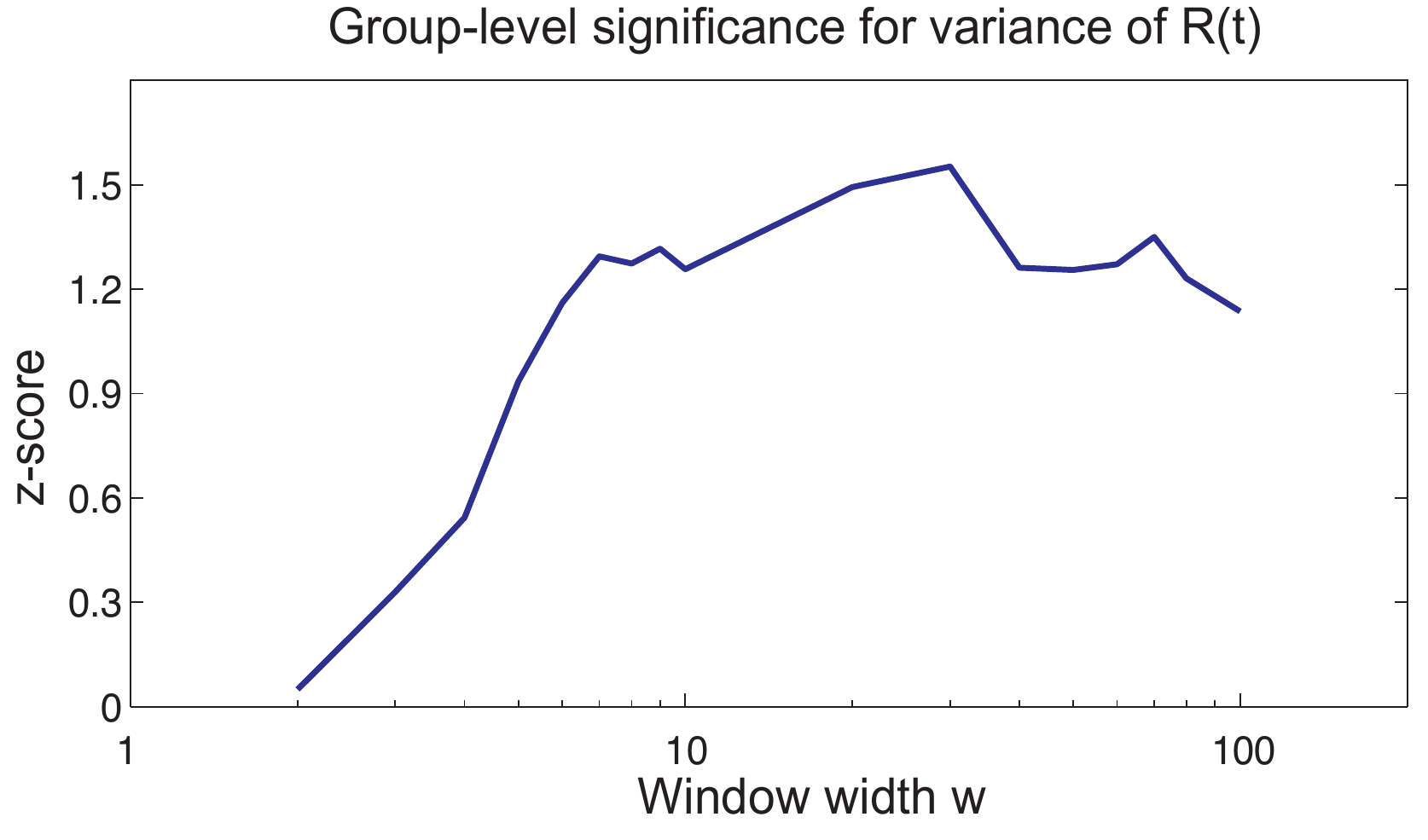}
	\caption{Estimation of the statistical significance region based on the variance of the dynamic correlation R(t)}
	\label{fig:varpv}
\end{figure}

In order to confirm the significance of the window width determined using V we used another marker to differentiate the original and surrogate datasets: the variance of the dynamic correlation R(t). The result shown in Figure \ref{fig:varpv} presents a peak a significance around w=30 TR, hereby confirming our findings.

\begin{figure}[h!]
	\centering
		\includegraphics[width=0.5\textwidth]{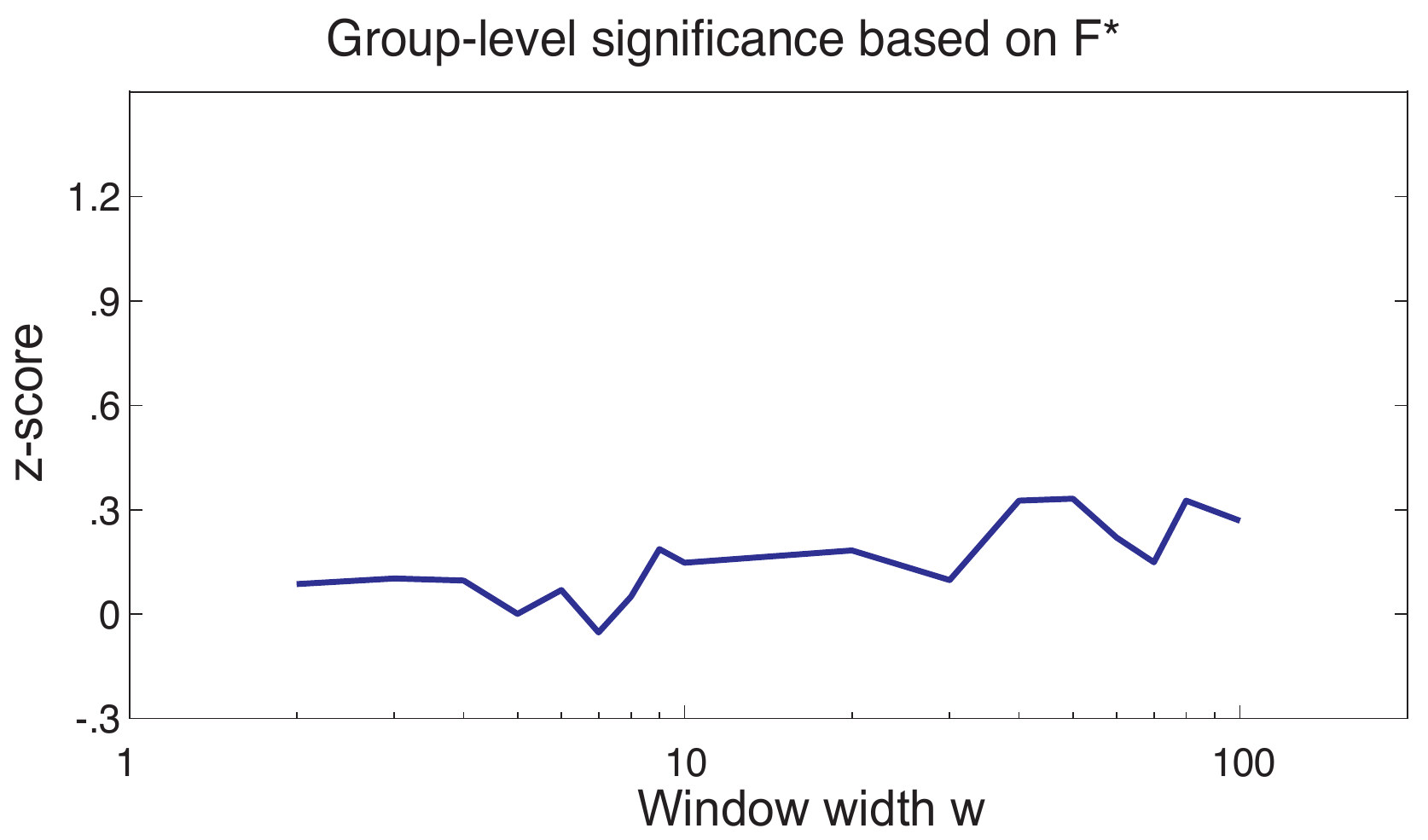}
	\caption{Estimation of the statistical significance region based on $F^*$}
	\label{fig:varf}
\end{figure}

Figure \ref{fig:varf} shows that it is not possible to differentiate original and surrogate datasets based on the main oscillatory mode of R(t). This suggests that $F^*$ is imposed by the window width $w$ and does not necessarily capture effects of neuronal dynamics.

\end{document}